\documentclass{article}
\usepackage[centertags]{amsmath}

\begin{document}

\title{Maximal nonsymmetric entropy leads naturally to Zipf's law }

\author{ Chengshi Liu \\Department of Mathematics\\Daqing Petroleum Institute\\Daqing 163318, China
\\Email: chengshiliu-68@126.com}

 \maketitle

\begin{abstract}
As the most fundamental empirical law, Zipf's law has been studied
from many aspects. But its meaning is still an open problem. Some
models have been constructed to explain Zipf's law. In the letter,
a new concept named nonsymmetric entropy was introduced,
maximizing nonsymmetric entropy leads naturally to
Zipf's law.


\end{abstract}

Zipf's law (\cite{Zi})which states that the frequency of a word
decays as a power law of its rank, is regarded as a basic
hypothesis with no need for explanation in recent models of the
evolution of syntactic communication(\cite{NW}). As an empirical
law, Zipf's law is the most fundamental fact in quantitative
linguistics and other complex systems, its meaning is still an
open problem which has been tried to explain from several aspects
of its origins(\cite{Ma,Si, NB,PT,CO}. It is necessary to find a
suitable mechanism for Zipf's law. In the present article, we
introduce a new concept named nonsymmetric entropy so that Zipf's
law is derived
naturally by maximizing the nonsymmetric entropy. \\

 We first give some hypothesis in a model of linguistics, in
which the set of words is $W=\{w_1,\cdots,w_m \}$, a text is a
sequence $s_L$ of words, that is, $s_L=\{s_1, \cdots, s_L\}$,
$s_i\in W, i=1, \cdots, L;$
 $N_i$ is the number of the word $w_i$ in a text
$s_L$, and $p_L(i)=\frac{N_i}{L}$ express the frequency of the
word $w_i$ in the text $s_L$; as $L \rightarrow\infty$, we assume
$p(i)=\lim\frac{N_i}{L}$, hence we have $p(1)+\cdots+p(m)=1$.
Furthermore if we assume $p(1)\geq p(2)\geq\cdots\geq p(m)$, then
 Zipf's law can be written as a power law in the form:
\begin{equation}\label{Zi}
p(i)=\frac{p(1)}{i^\alpha},
\end{equation}
where $\alpha\simeq1$ and $p(1)$ is the probability of the most
frequent word.\\

\textbf{Definition}: we define a function
\begin{equation}\label{WE}
S_m(p(1),\cdots,p(m))=-\sum _{i=1}^m p(i)\ln(\beta_ip(i)),
\end{equation}
where $\beta_i>0, i=1,\cdots, m$, are nonsymmetric parameters. We
call
the function $S_m$ the nonsymmetric entropy.\\

\textbf{Theorem}: If  $\{p(1),\cdots,p(m)\}$ satisfies the
following distribution
\begin{equation}
p(i)=\frac{1}{\beta_i\sum_{i=1}^m \frac{1}{\beta_i}},
\end{equation}
then the nonsymmetric entropy $S_m$ take the maximum
\begin{equation}
S_m=-\ln p(1)=\ln \sum_{i=1}^m \frac{1}{\beta_i}.
\end{equation}\\

Proof: Instituting $p(m)=1-p(1)-\cdots-p(m-1)$ into the
Eq.(\ref{WE}) and setting its differential to zero yields
\begin{equation}\label{Eu}
\frac{\partial S_m}{\partial
p(i)}=-\ln\frac{\beta_ip(i)}{\beta_m(1-p(1)-\cdots-p(m-1))}=0,
 \ \ i=1,\cdots,m-1,
\end{equation}
that is,
\begin{equation}\label{AE}
p(1)+\cdots+(1+\frac{\beta_i}{\beta_m})p(i)+\cdots+p(m-1)=1,
 \ \ i=1,\cdots,m-1.
\end{equation}
Solving the equations system (\ref{AE}), we obtain
\begin{equation}
p(i)=\frac{1}{\beta_i\Sigma_{j=1}^m\frac{1}{\beta_j}}=\frac{\beta_1}{\beta_i}p(1),
\ \i=1,\cdots,m,
\end{equation}
where $p(1)=\frac{1}{\beta_1\Sigma_{j=1}^m\frac{1}{\beta_j}}$.
Denote $A_k=(a_{ij})_{k\times k},
a_{ij}=\frac{\delta_{ij}}{p(i)}+\frac{1}{p(m)}$, since
$\det{A_k}=\frac{1}{p(m)}\Sigma_{i=1}^k\frac{1}{p(1)\cdots
\widehat{p(i)}\cdots p(k)}$, where the hat means the corresponding
item disappear, so from $a_{ij}=-\frac{\partial^2S_m}{\partial
p(i)\partial p(j)}=\frac{\delta_{ij}}{p(i)}+\frac{1}{p(m)}$, we
know that the matrix $A=(a_{ij})_{(m-1)\times(m-1)}$ is a positive
defined matrix. Thus the distribution
$p(i)=\frac{1}{\beta_i\sum_{i=1}^m \frac{1}{\beta_i}}$ maximize
the nonsymmetric entropy. The proof is completed.\\

\textbf{Corollary 1}. If we take $ \beta_i=i^\alpha$, then we have
\begin{equation}
p(i)=\frac{1}{\Sigma_{j=1}^m\frac{1}{j^\alpha}}=\frac{p(1)}{i^\alpha},
\ \i=1,\cdots,m,
\end{equation}
in particular, we take $\alpha\simeq1$, this is just the Zipf's
law. If take $\beta_i=(i+\gamma)^\alpha$, then we give Mandelbrot's law.\\

 \textbf{Corollary 2}. For Zipf's law, we have $S_{m+1}>S_m$, that is, the nonsymmetric entropy
 is increase as $m$ increasing.\\

 Proof: It is easy to see from $S_m=-\ln p(1)=\ln \sum_{i=1}^m
 \frac{1}{i}$.\\

The above results suggest that the nonsymmetric entropy is a more
fundamental concept that will play an important role in the fields
where Zipf's law emergences(\cite{Li} and references therein). The
meaning of the nonsymmetric entropy needs a reasonable
explanation. It is different with Shannon's entropy
$S=-\sum_{i=1}^m p(i) \ln p(i)$ in some aspects. For example, if
$p(j)=1$ and others zeroes, then $S=0$, but $S_m=-\ln\beta_j$,
this implies that there exist some kinds of uncertain in some
superficial reliable events under the nonsymmetric entropy. Other
deep meanings of nonsymmetric entropy need more studies. Since the
important roles of Boltzman's entropy and Shannon's entropy in
thermodynamics and information theory respectively, we tend to
take maximal nonsymmetric entropy principle but Zipf's law as a
basic principle.


\begin{thebibliography}{2}
\bibitem{Zi}Zipf, G. K. 1949 \emph{Human Behaviour and Principle of Least
Effort. An Introduction to Human Ecology}, Addison-Wesley,
Cambrighe, MA.
\bibitem{NW}Nowak, M. D., Plotkin, J. B. and Jansen, V. A. 2000
\emph{Nature} 404 495-498.
\bibitem{Ma}Mandelbrot, B. 1966 \emph{in Reading in Mathematical Social
Science}, eds. Lazarsfield, P. F and Henry, N. W. MIT Press,
Cambridge, MA, 151-168.
\bibitem{Si}Simon, H. A. 1955 \emph{Biomitrika}, 42 425-440.
\bibitem{NB}Naranan, S. and Balasubrahmanyan, V 1998 \emph{J. Quant.
Linguist}. 5, 35-61.
\bibitem{PT}Pietronero,L., Tosatti, E., Tosatti, V. and
Vespignani, A. 2001 \emph{Physica A} 293, 297-304.
\bibitem{CO}Ferrer i Cancho, R. and Sol\'{e}, R. V. 2003 \emph{PANS} 100,
788-791.
\bibitem{Li}Li, W. 2002 \emph{Glottometrics} 5 14-21
\end{thebibliography}
\end{document}